\documentclass[runningheads]{llncs}
\usepackage[T1]{fontenc}
\usepackage{graphicx}
\usepackage[pdftex]{hyperref}  

\usepackage{amssymb}
\usepackage{amsmath}

\usepackage{amsthm}
\usepackage{amsfonts}
\usepackage{cite}
\usepackage{algorithmic}
\usepackage{array}
\usepackage{float}
\usepackage{textcomp}
\usepackage{xcolor}
\usepackage{tabularx}
\usepackage{multirow}
\usepackage{booktabs} 
\usepackage{pifont} 
\usepackage[colorinlistoftodos]{todonotes} 
    
\newcommand{\cmark}{\ding{51}}%
\newcommand{\xmark}{\ding{55}}%

\usepackage{color}

\theoremstyle{definition}
\newtheorem{defn}{Definition}[]

\makeatletter
\def\th@definition{
  \thm@notefont{}
  \normalfont 
}
\makeatother

\begin{document}



\title{Why am I Waiting? Data-Driven Analysis of Waiting Times in Business Processes\thanks{Work funded by the European Research Council (PIX Project)}}

\titlerunning{Data-Driven Analysis of Waiting Times in Business Processes}

\author{Katsiaryna Lashkevich\orcidID{0000-0003-4426-7738} \and
Fredrik Milani\orcidID{0000-0002-1322-915X} \and
David Chapela-Campa\orcidID{0000-0002-4711-9653}  \and
Ihar Suvorau\orcidID{0000-0002-1862-2604}  \and
Marlon Dumas\orcidID{0000-0002-9247-7476}}
\authorrunning{K. Lashkevich et al.}

\institute{University of Tartu, Tartu, Estonia\\
\email{\{firstname.lastname\}@ut.ee}}
\maketitle              

\begin{abstract}
Waiting times in a business process often arise when a case transitions from one activity to another. Accordingly, analyzing the causes of waiting times of activity transitions can help to identify opportunities for reducing the cycle time of a process. This paper proposes a process mining approach to decompose the waiting time observed in each activity transition in a process into multiple direct causes and to analyze the impact of each identified cause on the cycle time efficiency of the process. An empirical evaluation shows that the proposed approach is able to discover different direct causes of waiting times. The applicability of the proposed approach is demonstrated on a real-life process.


\end{abstract}

\section{Introduction\label{sec:introduction}}


Waiting time is a common source of waste in business processes~\cite{delias2017positive}. 
Although it is impractical to completely eliminate waiting times in business processes, there are various approaches to reduce them~\cite{jansen2005business}. Waiting times typically arise during transitions between activities, i.e., when the execution of a case moves from one activity to another. There are different reasons why waiting times occur during activity transitions. For instance, when two consecutive activities in a case are executed by different resources (a.k.a.\ a handoff~\cite{van2007business}), the processing of the case is put on hold until the next resource becomes available to execute it. In this scenario, the cause of the waiting time is resource contention, i.e.\ a resource is not available to execute an activity instance because they are busy executing other activity instances~\cite{Dumas2013fundamentals}. Waiting times may also be caused by data exchanges~\cite{ramakrishnan2008defining}, coordination issues~\cite{rummel2005activity}, or synchronization points~\cite{rummel2005activity}.



Process mining techniques allow us to analyze data generated by business process executions, a.k.a.\ \emph{event logs}, to unveil performance and conformance issues, and associated improvement opportunities~\cite{Aalst2016}. 
In particular, process mining techniques support the discovery of sources of waste, including waiting times~\cite{Aalst2016}. 
However, while existing process mining techniques enable analysts to visualize activity transitions with high waiting time (i.e., bottlenecks), they provide limited support for identifying the causes of waiting times and how to reduce them.

To tackle this gap, this paper addresses the following research questions (RQ): (1) \textit{``What causes of waiting times between pairs of activity instances can be identified from event logs?''}, (2) \textit{``How can these causes of waiting time be identified from event logs?''}, and (3) \textit{``How can improvement opportunities, expressed as inefficiencies due to waiting times, be identified from event logs?''} 

The contribution of the paper is two-fold.
First, we conceptualize the causes of waiting time arising from activity transitions.
Second, we propose a process mining approach to (1) discover waiting times associated with activity transitions; (2) identify their causes; and (3) analyze their impact w.r.t.\ a well known measure of temporal efficiency called Cycle Time Efficiency (CTE): the ratio of processing time to cycle time in a process~\cite{Dumas2013fundamentals}. 


The proposed approach has been implemented as a software tool and  empirically evaluated to verify its ability to discover different causes of waiting time using synthetic event logs. In addition, the applicability of the approach has been tested by applying it to a real-life event log.

The rest of the paper is structured as follows.
Section~\ref{sec:related-work} introduces background and related work.
Section~\ref{sec:discovery-analysis} presents the proposed approach.
Section~\ref{sec:evaluation} describes the empirical evaluation, and Section~\ref{sec:conclusion} concludes the paper.

\section{Background and Related Work\label{sec:related-work}}

In this section, we introduce relevant conceptual foundations and notations from the fields of business process management and process mining, and position our contribution w.r.t.\ existing approaches to discover and analyze waiting times. 

\subsection{Business Processes and Temporal Performance Measures}

A business process is a collection of events, activities, and decisions that lead from a customer need to an outcome that is of value to this customer~\cite{Dumas2013fundamentals}. Each execution of a business process is called a \emph{case}.
A common measure of temporal performance of a business process is its \emph{cycle time}: The time between the moment a case of the process starts, and the case ends, aggregated to the level of the set of cases of a process observed during a period of time.
The cycle time of a process consists of \emph{processing time} (i.e., the time during which a case is being processed) and \emph{waiting time} (i.e., the time when a case waits to be processed)~\cite{Dumas2013fundamentals}. 




Waiting time may be caused by resource contention, i.e., no suitable resource is available to execute an activity instance~\cite{jansen2005business}. 
Other causes of waiting time include: synchronization between resources within a process~\cite{rummel2005activity} or across multiple processes~\cite{Dumas2013fundamentals}, coordination between resources executing different activities~\cite{rummel2005activity}, data transfer~\cite{Dumas2013fundamentals}, batching~\cite{lashkevich2022data}, handoffs~\cite{rummel2005activity}, and external inputs~\cite{AndrewsW17}. 


As stated earlier, the temporal efficiency of a process can be measured by its CTE: the ratio of processing time to cycle time.
When CTE is close to 1, there is relatively little waiting time. Conversely, if the CTE is close to 0, the waiting times are longer relative to processing times and there are opportunities to improve the CTE by reducing waiting times~\cite{Dumas2013fundamentals}.

\subsection{Event Logs and Activity Instance Logs}

Modern IT systems record and store process execution data in \emph{event logs}, i.e., sets of timestamped events capturing the execution of the activities in a process~\cite{Dumas2013fundamentals}.
An event log contains information about state changes of each activity instance (e.g.\ enablement, start, end, or cancellations of activity instances). In this paper, we use the concept of \emph{activity instance log} to represent the execution of a set of cases in a process.
An activity instance log is an event log in which each entry contains information about the start time and end time of an activity instance~\cite{Martin2021}. Below, we introduce several notations used in the paper, leading to a definition of an activity instance log.

We consider a business process that involves a set of \emph{activities} $A$.
We denote each of these activities with $\alpha$.
An \emph{activity instance} $\varepsilon = (\varphi, \alpha, \tau_e, \tau_s, \tau_c, \rho)$ denotes one execution of activity $\alpha$, where $\varphi$ identifies the case (represented by a \emph{process trace}) to which this execution belongs to, $\tau_e$, $\tau_s$, and $\tau_c$ denote, respectively, the instants in time in which this activity instance was \emph{enabled}, \emph{started}, and \emph{completed}, and $\rho$ identifies the \emph{resource} that processed the activity.
Accordingly, we use $\varphi(\varepsilon_{i})$, $\alpha(\varepsilon_{i})$, $\tau_e(\varepsilon_{i})$, $\tau_s(\varepsilon_{i})$, $\tau_c(\varepsilon_{i})$ and $\rho(\varepsilon_{i})$ to denote, respectively, the process trace, the activity, the enablement time, the start time, the completion time, and the resource associated with the activity instance $\varepsilon_{i}$. 
We use $(\tau_{i}, \tau_{j})$ to denote the interval of time starting in $\tau_{i}$ and ending in $\tau_{j}$.
In this way, $\omega(\varepsilon_{i}) = (\tau_e(\varepsilon_{i}), \tau_s(\varepsilon_{i}))$ denotes the \textit{waiting time} of $\varepsilon_{i}$, representing the interval since $\varepsilon_{i}$ became available for processing ($\tau_e(\varepsilon_{i})$), until its recorded start ($\tau_s(\varepsilon_{i})$).
Accordingly, the processing time of $\varepsilon_{i}$ is $pt(\varepsilon_{i}) = (\tau_s(\varepsilon_{i}), \tau_c(\varepsilon_{i}))$.
We use $(\tau_{i},\tau_{j}) \in (\tau_{k},\tau_{l})$ to denote that the interval $(\tau_{i},\tau_{j})$ is contained in $(\tau_{k},\tau_{l})$, i.e.\ $\tau_{i} \geq \tau_{k}$ and $\tau_{j} \leq \tau_{l}$.
With $(\tau_{i},\tau_{j}) \perp (\tau_{k},\tau_{l})$ we denote that both intervals (partially or fully) overlap, i.e., $\exists (\tau_{m},\tau_{n}) \in (\tau_{i},\tau_{j}) \mid (\tau_{m},\tau_{n}) \in (\tau_{k},\tau_{l})$.

Given the above, an \textit{activity instance log} $L$ is a collection of activity instances recording the data of the execution of a set of traces of a business process.

\subsection{Related Work}
Process mining is a research field that encompasses a wide range of techniques that enable process discovery and efficiency, quality, compliance, predictive, and prescriptive analysis \cite{milani2022process}.
Some techniques address the question of how to discover and analyze waiting times in specific application domains. For instance, Uysal et al.~\cite{uysal2020process} present a case study where process mining is used to identify bottlenecks and reduce the cycle time in a production process.
Similarly, Erdogan et al.~\cite{erdogan2022multi} apply process mining techniques to identify waiting times in a hospital emergency process, while Yampaka \& Chongstitvatana~\cite{yampaka2016application} describe an application of process mining combined with a queuing system to analyze and improve temporal performance in a healthcare process. Similarly, Antunes et al.~\cite{antunes2019solution} combine process mining with discrete event simulation to optimize waiting time in an emergency department. 
However, none of these domain-specific studies considers the question of how to attribute waiting times to their causes, which is the focus of this paper.

Ferreira \& Vasilyev~\cite{ferreira2015using} present a technique to identify why some cases in a process take longer time to complete. 
They identify case characteristics correlating with higher delays, e.g., when a given activity occurs in a case, or when a given resource is involved, the case is likely to have higher waiting time. Likewise,
De Leoni \& van der Aalst~\cite{de2016general} combine some of the existing correlation analysis techniques to identify how different process characteristics correlate with the process performance, e.g., if process deviations cause delays.
Similarly, Hompes et al.~\cite{hompes2017discovering} propose an approach based on time series analysis to detect cause\mbox{-}effect relations between process characteristics and performance indicators, e.g., if the waiting time for the receipt of a payment depends on the time of day.
Toosinezhad et al.~\cite{toosinezhad2020detecting} introduce an approach to detect event patterns that frequently precede, i.e. lead to, dynamic bottlenecks.
While these studies take a correlation-based approach to analyze waiting times, we classify the causes of waiting times and consider their impact on process performance. 

Some process mining techniques support the identification of waiting times caused by queuing effects, i.e., when activity instances wait in a queue until a resource becomes available~\cite{senderovich2015queue}. 
Similarly, in~\cite{lashkevich2022data}, the authors present an approach to discover waiting times caused by batch processing.
However, these techniques focus on identifying a singular cause of waiting time. In contrast, in the present paper we seek to decompose the observed waiting time into multiple causes.

\section{Waiting Time Discovery and Analysis\label{sec:discovery-analysis}}

\begin{figure*}[b]
    \centering
    \includegraphics[width=0.99\linewidth]{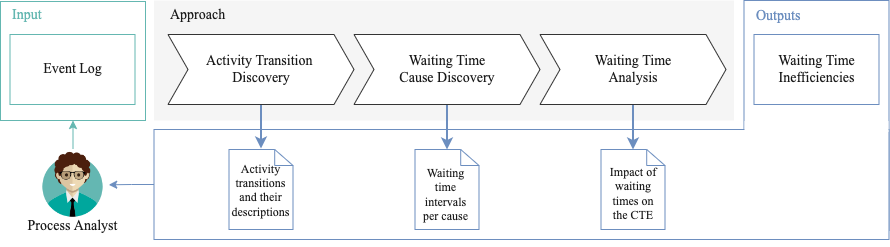}
    \caption{Overview of the proposed approach.}\par
    \label{fig:mesh1}
\end{figure*}

In this section, we describe our approach to discover and analyze waiting times in a business process.
The approach takes an event log as input and, as a result, produces a report comprising the causes of waiting times in a business process and their impact on the CTE.
\figurename~\ref{fig:mesh1} depicts an overview of the three main steps of the proposed approach.

In the first step, we discover the transitions between activities and their characteristics -- total frequency, case frequency, and total waiting time -- from the event log.
In the second step, we identify the causes of waiting time of each transition. 
In the third and final step, we analyze the impact of each cause on the temporal efficiency of the process.

\subsection{Activity Transition Discovery}

\begin{figure}[h]
    \centering
    \includegraphics[width=0.95\linewidth]{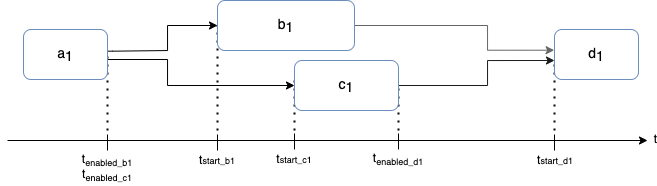}
    \caption{Waiting time in a case fragment with concurrent activity instances.}\par
    \label{fig:mesh2}
\end{figure}

The first step of our approach is to discover transitions between activities and their waiting times.
To do this, we define an \textit{activity transition instance} as a pair of activity instances $\langle a_{1}, b_{1} \rangle$ in a single case, such that the completion of $a_{1}$ enables $b_{1}$, i.e., $b_{1}$ cannot be executed before $a_{1}$ is completed.
In a sequential process, each activity instance of a case is enabled by the completion of the preceding activity instance.
However, concurrency is common in real-life processes.
\figurename~\ref{fig:mesh2} shows an example of the execution of a sub-process formed by four activities with concurrency between two of them.
In this case, even though the order of the activity instances is $a_1$, $b_1$, $c_1$, and $d_1$, both $b_1$ and $c_1$ are enabled when the activity instance $a_1$ completes.
In the same way, the activity instance $d_1$ is enabled only when $c_1$ is completed.
Thus, there are only three activity transitions in this example, namely $\langle a_1, b_1 \rangle$, $\langle a_1, c_1 \rangle$, and $\langle c_1, d_1 \rangle$.

To detect the concurrency relations, e.g., when the activities $b$ and $c$ are concurrent (see \figurename~\ref{fig:mesh2}), we use the concurrency oracle from the Heuristics Miner~\cite{DBLP:conf/cidm/WeijtersR11}.
This method computes, for each observed relation between two activities, a degree of confidence whether it is a concurrent or directly-follows relation, based on the percentage of occurrences in each order -- $b$ followed by $c$, or $c$ followed by $b$.
Then, based on a set of defined thresholds, it retrieves the concurrent relations of the process.
In this way, we consider each activity instance being enabled by its closest non-concurrent preceding activity instance.
We assume that once an activity instance is enabled, it remains enabled until its processing starts. 

We call the first element of an activity transition instance the \emph{source activity instance}, while the latter is the \emph{target activity instance}.
An \emph{activity transition} is a set of activity transition instances with the same source and target activities, where the \emph{source activity} is the activity executed in all its source activity instances, and the \emph{target activity} is the activity executed in all its target activity instances.
For example, the activity transition $\langle a, b \rangle$ is composed of the set of activity transition instances $\{\langle a_1, b_1 \rangle, \langle a_2, b_2 \rangle, ..., \langle a_n, b_n \rangle\}$ in the event log.
For the sake of simplicity, we will refer to activity transitions as \emph{transitions}, and to activity transition instances as \emph{transition instances}.

Once the transition instances are discovered, we calculate their \emph{duration}, i.e., the waiting times they induce.
The waiting time of a target activity instance in a transition instance is the interval between its enablement and its start time.
For example, in \figurename~\ref{fig:mesh2}, the waiting time in the transition $\langle c_1, d_1 \rangle$ corresponds to the interval $(t_{enabled\_d1}, t_{start\_d1})$. 
In this way, we identify the waiting time of each target activity instance per transition.

Finally, for each identified transition (composed of all its transition instances), we compute the following characteristics. 
\emph{Case frequency} illustrates the proportion of process cases from total number of cases where this transition is observed.
\emph{Total frequency} indicates the number of occurrences of this transition in the process. 
\emph{Total duration} is the sum of the waiting times of all transition instances.
The output of this step is a report depicting all identified transitions and their characteristics, sorted by total duration in descending order.
Based on this information, the analysts can see what transitions cause the highest waiting times and how frequently they are executed.

\subsection{Waiting Time Cause Discovery\label{subsec:wt-type-discovery}}

Once the activity transitions and their characteristics are discovered, we analyze the waiting time of each transition instance and identify their causes.
We target five causes of waiting time (WT): WT due to batching, resource contention, prioritization, resource unavailability, and extraneous factors.
In this section, we define these five causes of waiting times (RQ1) and describe how they can be identified from an event log (RQ2).

\begin{figure}[t]
    \centering
    \includegraphics[width=\linewidth]{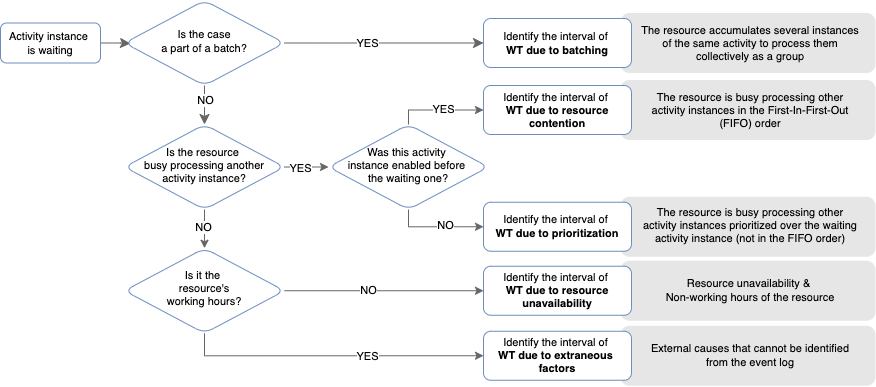}
    \caption{Overview of the waiting time cause discovery process and their definitions.}\par
    \label{fig:mesh3}
\end{figure}

Within a single transition instance, waiting time can be explained by one or several causes --e.g.\ the resource was working in another activity half of the waiting time, and out of their working hours the other half.
If there are several waiting time causes in a transition, we propose to classify them in non-overlapping time intervals by identifying them in strict order using the dichotomies' classification illustrated in \figurename~\ref{fig:mesh3}.
Thus, when an interval of the waiting time is classified as a certain cause, it cannot be classified as any other.

We first identify if any intervals of the transition duration are caused by batching. 
Then we seek intervals of waiting time caused by resource contention and prioritization, followed by resource unavailability and extraneous factors (see \figurename~\ref{fig:mesh3}).
The order of identification is based on the dominance of a certain cause over the others, i.e., although several phenomena might be observed for the same interval, only the cause with the highest dominance is considered. 
For instance, if during the same waiting time interval, both batching and resource unavailability exist, we consider the primary cause for the instance waiting, i.e., batching in this case.
Although the resource might be unavailable, the processing cannot start until the batch is accumulated. 
Therefore, when multiple causes are observed, only one is considered as the direct cause of waiting time. 
In addition, this approach ensures the addition property of the waiting time for each transition instance -- i.e.\ the sum of the waiting time causes is equal to the total waiting time of the transition instance.

\vspace{10pt}

\noindent\textbf{Waiting time due to Batching.} The first cause that we identify is \emph{waiting time due to batching}.
Batch processing occurs when a set of instances of the same activity are accumulated to be processed together (either simultaneously or one after the other)~\cite{lashkevich2022data}.
In this context, a batch is a set of activity instances $\mathcal{E}_b \subseteq L$, such that all $\varepsilon_{i} \in \mathcal{E}_b$ record the execution of the same activity, performed by the same resource, and processed as a batch (i.e.\ all of them were enabled before any of them started, and they were processed as a group).
We use the technique proposed in~\cite{lashkevich2022data} to identify batch processing.
In batch processing, when an activity instance is enabled and ready to be processed, it can wait for other instances of the same activity until the batch is accumulated, i.e., until all instances that are a part of a batch are collected.

\begin{figure}[t]
    \centering
    \includegraphics[scale=0.5]{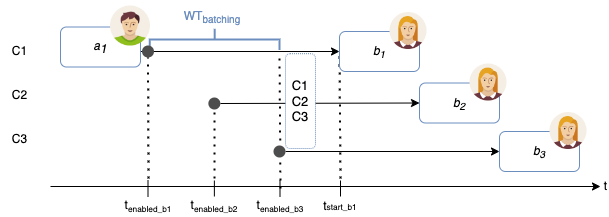}
    \caption{Waiting time due to batching.}\par
    \label{fig:mesh4}
\end{figure}

Accordingly, when the target activity instance of a transition instance is detected as part of a batch, the waiting time interval from its enablement time to the batch accumulation time is classified as \emph{waiting time due to batching} (see \figurename~\ref{fig:mesh4}).
The waiting time due to batching of an activity instance is then defined as follows:

\begin{defn}[Waiting Time Due to Batching\label{def:batching-wt}]

    Given an activity instance log $L$, a batch $\mathcal{E}_b \subseteq L$, and an activity instance $\varepsilon_{i} \in \mathcal{E}_b$. The WT due to batching of $\varepsilon_{i}$ is $\omega_{ba}(\varepsilon_{i}) = (\tau_{e}(\varepsilon_{i}), \tau_{bc}) \mid \tau_{bc} = max(\{\tau_{e}(\varepsilon_{j}) \mid \varepsilon_{j} \in \mathcal{E}_b\})$, i.e.\ the interval of time between the enablement of $\varepsilon_{i}$ and the last enablement of the activities in the batch.
    
\end{defn}

Following with the example in \figurename~\ref{fig:mesh4}, the activity $B$ is a batch processed activity where the resource accumulates instances and then processes them one by one (sequential batch processing)~\cite{Martin2021}.
The batch is accumulated until the activity instance $b_3$ is enabled (case C3 is ready to be processed ($t_{enabled\_b3}$). 
Therefore, if we analyze the transition instance $\langle a_1, b_1 \rangle$ of case \texttt{C1}, its WT due to batching corresponds to the interval of the waiting time between $t_{enabled\_b1}$ and $t_{enabled\_b3}$.

\vspace{10pt}

\begin{figure}[t]
    \centering
    \includegraphics[width=\linewidth]{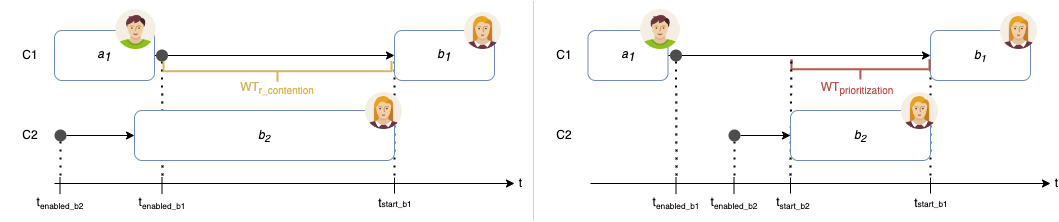}
    \caption{Waiting time due to resource contention and due to prioritization.}\par
    \label{fig:mesh5}
\end{figure}

\noindent\textbf{Waiting time due to resource contention.} 
There are situations where the resources that have to process a certain activity are busy processing another activity instances that were enabled earlier than the waiting one and, thus, it's understood that they start processing them before the current one (following a first-in-first-out order).\footnote{
We assume that resources work only on one activity at a time, i.e., there is no multitasking.
Thus, we foresee that the proposed estimation technique will not be suitable for event logs with a high proportion of multitask activity instances.
}
When this situation occurs, we classify as \emph{waiting time due to resource contention} those intervals in which the resource that performed the activity instance was working in other activity instances enabled before it (see \figurename~\ref{fig:mesh5}).
Therefore, the waiting time due to resource contention of an activity instance is defined as follows:


\begin{defn}[Waiting Time Due to Resource Contention\label{def:contention-wt}]

    Given an activity instance log $L$, and an activity instance $\varepsilon_{i} \in L$. The WT due to resource contention of $\varepsilon_{i}$ is $\Omega_{rc}(\varepsilon_{i}) = 
    \{
    (\tau_{i}, \tau_{j}) \mid \
      \tau_{i} = max(\tau_e(\varepsilon_{i}), \tau_s(\varepsilon_{j})) \wedge
      \tau_{j} = min(\tau_s(\varepsilon_{i}), \tau_c(\varepsilon_{j})) \wedge
      \varepsilon_{j} \in L \wedge
      \varepsilon_{j} \neq \varepsilon_{i} \wedge
      \rho(\varepsilon_{j}) = \rho(\varepsilon_{i}) \wedge
      \tau_e(\varepsilon_{j}) \leq \tau_e(\varepsilon_{i}) \wedge
      pt(\varepsilon_{j}) \perp \omega(\varepsilon_{i})
    \}
    $, 
    i.e., the set of intervals of processing time of all $\varepsilon_{j}$ of $L$ (executed by the same resource as $\varepsilon_{i}$, and enabled before it) overlapping with the WT of $\varepsilon_{i}$. 

\end{defn}




\vspace{5pt}

\noindent\textbf{Waiting time due to prioritization.}
However, the resources might not always follow the FIFO policy.
In some situations, the resources might give priority to certain activity instances over others.
We call this behavior \emph{prioritization}, meaning that an activity instance is processed out of turn w.r.t. a FIFO policy, thus causing other activity instances to wait longer.
When this situation occurs, we classify as \emph{waiting time due to prioritization} those intervals in which the resource that performed the activity instance was working in other activity instances enabled after it (see \figurename~\ref{fig:mesh5}).
Therefore, the waiting time due to prioritization of an activity instance is defined as follows:



\begin{defn}[Waiting Time Due to Prioritization\label{def:prioritization-wt}]

    Given an activity instance log $L$, and an activity instance $\varepsilon_{i} \in L$. The WT due to prioritization of $\varepsilon_{i}$ is $\Omega_{prior}(\varepsilon_{i}) = 
    \{
    (\tau_{i}, \tau_{j}) \mid \
      \tau_{i} = \tau_s(\varepsilon_{j}) \wedge
      \tau_{j} = min(\tau_s(\varepsilon_{i}), \tau_c(\varepsilon_{j})) \wedge
      \varepsilon_{j} \in L\ \wedge\ 
      \varepsilon_{j} \neq \varepsilon_{i} \wedge
      \rho(\varepsilon_{j}) = \rho(\varepsilon_{i})\ \wedge\ 
      \tau_e(\varepsilon_{j}) > \tau_e(\varepsilon_{i})\ \wedge\ 
      pt(\varepsilon_{j}) \perp \omega(\varepsilon_{i})
    \}
    $, 
    i.e., the set of intervals of processing time of all $\varepsilon_{j}$ of $L$ (executed by the same resource as $\varepsilon_{i}$, and enabled after it) overlapping with the WT of $\varepsilon_{i}$. 

\end{defn}


\vspace{5pt}

\noindent\textbf{Waiting time due to resource unavailability.}
The fourth cause of waiting time that we propose to consider is resource unavailability, which corresponds to the intervals in time in which the resource is not available to work due to their working schedules.
To identify this waiting time, we need to first discover the working schedules of the resources.
We propose to use the technique presented in~\cite{DBLP:conf/bpm/Lopez-PintadoD22} to discover calendars over time granules not fully described by the input data.
This technique analyzes the instants in time when each resource interacted with the system (i.e.\ the start and end of each activity instance) to build a weekly working calendar composed of time intervals in which there was enough evidence, based on given support and confidence values, that the resource is working.\footnote{Although we use this resource calendar discovery algorithm, the approach can be applied with other calendar discovery algorithms or with manually defined calendars.}
Given the weekly calendars of each resource, we transform them to absolute time intervals to compare them with the waiting times observed in the log.
Then, we classify as \emph{waiting time due to resource unavailability} those intervals where the resource is not available for working.
Therefore, the waiting time due to resource unavailability of an activity instance is defined as follows:


\begin{defn}[Waiting Time Due to Resource Unavailability\label{def:unavailability-wt}]

    Given an activity instance log $L$, an activity instance $\varepsilon_{i} \in L$, and being $cal_{av}(\rho) = \{(\tau_{avs}, \tau_{avc})\}$ a resource availability calendar with the set of time intervals in which the resource is available to work.
    The WT due to resource unavailability of $\varepsilon_{i}$ is  
    $\Omega_{unav}(\varepsilon_{i}) = 
    \{
    (\tau_{i}, \tau_{j}) \mid \
      (\tau_{i}, \tau_{j}) \in \omega(\varepsilon_{i}) \wedge
      \nexists (\tau_{k}, \tau_{l}) \in cal_{av}(\rho(\varepsilon_{i})) \mid (\tau_{i},\tau_{j})
      \perp (\tau_{k},\tau_{l})
    \}
    $,
    i.e., the set of intervals of the waiting time of $\varepsilon_{i}$ that do not overlap with the availability calendar of the resource $\rho(\varepsilon_{i})$ that executed $\varepsilon_{i}$.
    
\end{defn}

\begin{figure}[t]
    \centering
    \includegraphics[scale=0.5]{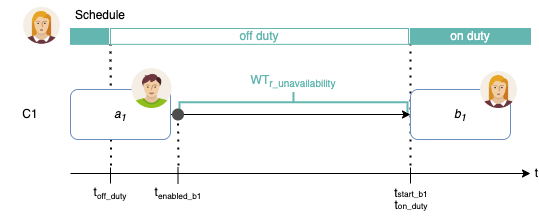}
    \caption{Waiting time due to resource unavailability.}\par
    \label{fig:mesh7}
\end{figure}

\vspace{5pt}

\noindent\textbf{Waiting time due to extraneous factors.}
The last cause of waiting time that we propose to consider is ``extraneous factors''.
We propose to classify as \emph{waiting time due to extraneous factors} the waiting time intervals caused by the external effects that cannot be identified from the event log -- e.g.\ the resource is working on another process, the activity instance cannot start until some unrecorded event has occurred, fatigue effects, or context switch.
Thus, the waiting time of an activity instance due to extraneous factors is defined as follows.


\begin{defn}[Waiting Time Due to Extraneous Factors\label{def:extraneous-wt}]

    Given an activity instance log $L$ and an activity instance $\varepsilon_{i} \in L$.
    The WT due to extraneous factors of $\varepsilon_{i}$ is 
    $\Omega_{extr}(\varepsilon_{i}) = 
    \{
        (\tau_{i},\tau_{j}) \mid \
        (\tau_{i},\tau_{j}) \in \omega(\varepsilon_{i}) \wedge
        \nexists (\tau_{k},\tau_{l}) \in (
            \{\omega_{ba}(\varepsilon_{i})\} \bigcup 
            \Omega_{rc}(\varepsilon_{i}) \bigcup
            \Omega_{prior}(\varepsilon_{i}) \bigcup
            \Omega_{unav}(\varepsilon_{i})
        ) \mid (\tau_{i},\tau_{j}) \perp (\tau_{k},\tau_{l}) 
    \}$,
    i.e., the set of intervals within the WT of $\varepsilon_{i}$ that does not overlap with waiting times due to batching, resource contention, prioritization, or resource unavailability.
    
\end{defn}

Therefore, the second step of the approach results in a report discovering and measuring the total waiting time for each transition according to their causes. 
These results reveal the cause of each waiting times and by how much. 

\subsection{Waiting Time Analysis}

The final step is to analyze how much each waiting time cause contributes to the temporal performance of the process and their impact to the CTE.
With that purpose, we propose to compute the percentage of time that each cause induces in the CTE of the process.
In this context, we measure the CTE as the processing time (PT) divided by the sum of the processing and the waiting time (PT $+$ WT), where PT is the sum of the processing time of all activity instances in the process, and WT is the sum of their waiting time.
The impact of waiting times per cause is calculated as the difference between the original process CTE and the CTE if a particular waiting time is eliminated.
In this way, we can measure (1) the impact that each waiting time cause has on the process CTE, (2) the impact that each transition has on the process CTE, and (3) the impact that each waiting time cause has in each transition.
This metrics can indicate the potential CTE improvement if a particular cause of waiting time is addressed.

As a result of the proposed analysis, we can analyze the discovered waiting time causes and their impact on CTE.
With this information, process analysts can identify where the inefficiencies due to waiting times are localized, choose which transitions and/or waiting time causes to address, and which redesign alternatives to apply.

\section{Evaluation\label{sec:evaluation}}

In this section, we present the evaluation of our approach.
We evaluate the approach by addressing two evaluation questions (EQ): (1) \textit{To which extent is the technique able to detect the presence or absence of certain waiting time causes?}, (2) \textit{To what extent is the technique able to correctly quantify the amount of waiting time waste per each cause?} 
In this experimentation we use synthetic data to validate the ability of the technique to accurately discover transitions, their waiting times, and the waiting time causes known to be present in the event logs. 
Then, we demonstrate the approach's applicability on a real-life event log.
The approach implementation, the event logs and experiment results, are all available on GitHub.\footnote{\url{https://github.com/AutomatedProcessImprovement/waiting-time-analysis/tree/caise2023}}

\subsection{Evaluation on Synthetic Data}

To answer EQ1, we used a business process simulation model (BPS model) of a loan application process to simulate a set of event logs with different combinations of waiting time causes.
To simulate WT due to resource contention, we set a low number of available resources in the BPS model, so that in some cases there were no resources available to process an enabled activity instance.
To create WT due to resource unavailability, we set resource working calendars so that some resources worked from Monday to Wednesday, and others from Thursday to Friday.
To simulate WT due to extraneous factors, we added timer events before some of the activities in the BPS model, thus delaying their start.
WT due to batching and prioritization cannot be injected by modifying the simulation parameters, as current BPS engines do not support them.
Therefore, in case of batching, we added a set of new traces delaying the start of some activity instances so that they are processed as a batch.
To simulate prioritization, we added a set of new traces changing the order of execution of some activity instances, so they are processed following a prioritization order.
Combining these modifications, we simulated a set of 32 event logs with all the combinations of causes of waiting time
and measured the performance using precision and recall.
True positives and false positives stand for the discovery of a waiting time cause that, respectively, was and was not injected in the event log.
True and false negatives denote an undiscovered waiting time cause that, respectively, was not and was injected.

\begin{table*}[t]
    \centering
    \scriptsize
    \caption{Results for the simulated event logs with all waiting time causes, depicting the true positives with '\cmark`, the false positives with '\xmark`, and the true negatives with an empty cell (there are no false negatives).
    }
    \label{tab:simulated-results}
    \resizebox{\textwidth}{!}{
    \begin{tabular}{l c c c c c c c c c c c c c c c c}
                            & S01    & S02    & S03    & S04    & S05    & S06    & S07    & S08    & S09    & S10    & S11    & S12    & S13    & S14    & S15    & S16    \\ \toprule
        Batching            &        & \cmark &        &        &        &        & \cmark & \cmark & \cmark & \cmark &        &        &        & \xmark &        &        \\
        Prioritization      &        &        & \cmark &        &        &        & \cmark &        &        &        & \cmark & \cmark & \cmark & \xmark & \xmark &        \\
        Res. Contention     &        &        &        & \cmark &        &        &        & \cmark &        &        & \cmark &        &        & \cmark & \cmark &        \\
        Res. Unavailability &        &        &        &        & \cmark & \xmark &        &        & \cmark & \xmark &        & \cmark & \xmark & \cmark & \xmark & \cmark \\
        Extraneous factors          &        &        &        &        &        & \cmark &        &        &        & \cmark &        &        & \cmark &        & \cmark & \cmark \\ \midrule
                            & S17    & S18    & S19    & S20    & S21    & S22    & S23    & S24    & S25    & S26    & S27    & S28    & S29    & S30    & S31    & S32    \\ \toprule
        Batching            & \cmark & \cmark & \cmark & \cmark & \cmark & \cmark & \xmark &        &        & \xmark & \cmark & \cmark & \cmark & \cmark & \xmark & \cmark \\
        Prioritization      & \cmark & \cmark & \cmark &        & \xmark &        & \cmark & \cmark & \cmark & \xmark & \cmark & \cmark & \cmark & \xmark & \cmark & \cmark \\
        Res. Contention     & \cmark &        &        & \cmark & \cmark &        & \cmark & \cmark &        & \cmark & \cmark & \cmark &        & \cmark & \cmark & \cmark \\
        Res. Unavailability &        & \cmark & \xmark & \cmark & \xmark & \cmark & \cmark & \xmark & \cmark & \cmark & \cmark & \xmark & \cmark & \cmark & \cmark & \cmark \\
        Extraneous factors          &        &        & \cmark &        & \cmark & \cmark &        & \cmark & \cmark & \cmark &        & \cmark & \cmark & \cmark & \cmark & \cmark \\ \bottomrule
    \end{tabular}
    }
\end{table*}

\tablename~\ref{tab:simulated-results} depicts the results for the simulated event logs used to evaluate EQ1.
Our approach discovered the injected waiting time causes in all the event logs (no false negatives) resulting in a recall of 100\%.
However, due to the influence that some waiting time causes have between them, the results contain 17 false positives (precision of 83\%).
False positives in WT due to resource unavailability (S06, S10, S13, S15, S19, S21, S24, and S28) are caused by the presence of extraneous waiting time, combined with limited data for the discovery of resources' working calendars.
To simulate these logs, we set 24/7 working calendar for all resources (high availability).
However, when a resource has low occupation (executes few events), there is not enough data for the calendar discovery to identify a 24/7 calendar and some intervals are interpreted as non-working time.
When these non-working intervals overlap with extraneous waiting time, the tool classifies them as WT due to resource unavailability.
This limitation is inherent to the discovery of the resource calendar, if there is no data showing that a resource was active during a period of time, it cannot be assumed that they were working.

The injection of WT due to extraneous factors also induced false positives of prioritization (S15, S21, S26, S30).
When an activity instance is enabled but waiting due to extraneous factors, the resource might execute other activities enabled after the waiting one, being detected as WT due to prioritization.
These false positives are due to the absence of an explicit indicator of the WT due to extraneous factors (i.e.\ extraneous waiting time is only detected when no other causes are identified).

False positives due to batching (S14, S23, S26, S31) are caused by the appearance of a batch processing behavior that was not intentionally added.
To create WT due to resource contention and unavailability, in some scenarios we assigned working calendars that produced low resource availability. 
In such cases, while instances were waiting until the resource became available, they were collected and then, executed by the same resource. 
This resulted in an unassigned but correctly discovered batch processing behavior.


EQ2 aimed at assessing if our approach is able to correctly identify waiting time intervals and their causes. 
We could not use the set of event logs used for EQ1 to answer EQ2, as the logs were created through stochastic business process simulation -- we know we have introduced a certain cause of waiting time, but we don't know to what extent. 
Therefore, 
we manually created a set of 5 event logs (with up to 15 traces) with activity transitions having different waiting time causes.
Due to the low number of events per resource, we manually defined the resource working calendars for this experiment.
We ran our technique over these logs and obtained accurate waiting time intervals and their causes.

\subsection{Evaluation on a Real-Life Log}

To evaluate the applicability of the proposed approach in a real-life scenario, we used an event log of a manufacturing production process~\cite{dafnalevy_2014}.
The event log has 225 traces (cases), 
recording the execution of 24 activities in a total of 4,503 activity instances,
executed by 46 resources.

First, we discovered activity transition instances and their characteristics: 91 transitions with 3421 transition instances (i.e., executions of each transition).
For each transition, we identified its characteristics (case frequency, total frequency and total duration, i.e., total waiting time).

Second, we discovered the causes of waiting times.
As a result, we obtained a report that captures how much of the waiting time is induced by each cause in every activity transition. 
The highest waiting times originated from the self-loop transitions, i.e., when the same activity is executed for the case twice in a row. 
For instance, transitions between two activity instances of ``Turning \& Milling Q.C.'' induced a total waiting time of 1101d 5h 34m. This made it the greatest contributor to the total waiting time in the process. 
The largest portion of the waiting time in this transition (54.74\%) was caused by resource unavailability and equaled 602d 18h 59m.



Finally, we analyzed the waiting times by calculating how much each cause contributed to the total WT of the process and how they affected the CTE.
Thus, resource unavailability was the major source of WT as it caused 57\% of the total WT. 
It is followed by batching (22\%), prioritization (9\%), extraneous (8\%) and resource contention (4\%).
Then, we measured the impact of the waiting times by cause on the process CTE (CTE = 6.81\%).
The highest CTE increase (up to 14.62\%) can be achieved if the WT due to resource unavailability is eliminated.
From the transition perspective, the highest improvement opportunity lies in addressing the ``Turning \& Milling Q.C'' self-loop transition. 
If the WT is this transition is addressed, the CTE could increase to 7.69\%.

\section{Conclusion\label{sec:conclusion}}

This paper outlines a process mining-based approach for identifying causes of waiting times and their impact.
To address research question RQ1, we propose a method to attribute the waiting time of activity transitions to one of five causes: \textit{batching}, \textit{resource contention}, \textit{prioritization}, \textit{resource unavailability}, and \textit{extraneous factors}.
To address RQ2, we outline how these five causes of waiting time can be discovered, for each activity transition, from an event log. 
Finally, we propose to measure the impact of each waiting time cause on the CTE of a process, to help analysts to prioritize opportunities to reduce these waiting times (RQ3). 
The empirical evaluation shows that our approach can accurately classify the waiting times in a process into the five causes, in (synthetic) event logs where the causes of waiting time are known.
Finally, we illustrated the applicability of our approach using an event log of a production process.


In its current form, the proposed approach only considers waiting times associated with transitions between activity instances. Yet waiting times may also arise in at least two other settings: (i) between case creation and start of the first activity instance; and (ii) within an activity instance due to interruptions (e.g. the resource interrupts their work and resumes it later). The first of these  waiting times could be analyzed by applying methods that estimate the inter-arrival time of each trace~\cite{DBLP:conf/bpm/MartinDC15,DBLP:conf/icpm/BerkenstadtGSSW20}. The second approach requires new methods for modeling and inferring interruptions, possibly using additional attributes in an event log or other additional data. 

Another limitation of the approach is that it does not consider multitasking. This could be addressed by inferring multitasking patterns from the log, and using this information to estimate at what point in time a resource would normally have started an activity instance, given their past multitasking behavior. 




In future work, we plan to develop a method for discovering business process simulation models from event logs, which takes into account the causes of waiting times considered in this paper. Such simulation models could be used to support analysts in identifying combinations of redesign options to optimize CTE, while also considering other performance dimensions (e.g.\ cost).




\bibliographystyle{splncs04}
\bibliography{bibliography}
    
\end{document}